\documentclass[11pt]{article}
\usepackage{latexsym}  % for Symbol fonts
\usepackage{amssymb}
\usepackage{graphicx}
\usepackage{amsmath}
\newcommand{\be}{\begin{equation}}
\newcommand{\ee}{\end{equation}}
\newcommand{\bea}{\begin{eqnarray}}
\newcommand{\eea}{\end{eqnarray}}
\newcommand{\bref}[1]{(\ref{#1})}
\begin{document}
%%%%%%%%%%%%%%%%%%%%%%%%%%%%
\begin{titlepage}
%%%%%%%%%%%%%% PREPRINT NUMBERS %%%%%%%%%%%%
\begin{flushright}
%\today
\end{flushright}
%%%%%%%%%%%%%%%%%%%%%%%%%%%%%%
\vspace{4\baselineskip}
%%%%%%%%%%%%%%%%%%% TITLE %%%%%%%%%%%%%%%%%%
\begin{center}
{\Large\bf  Measuring the lower bound of the neutrino mass at the LHC in Higgs triplet model}
\end{center}
%%%%%%%%%%%%%%%% AUTHORS %%%%%%%%%%%%%%%%%%%
\vspace{1cm}
\begin{center}
{\large Hiroyuki Nishiura$^{a,}$
\footnote{E-mail:nishiura@is.oit.ac.jp}}
and
{\large Takeshi Fukuyama$^{b,}$
\footnote{E-mail:fukuyama@se.ritsumei.ac.jp}}
\end{center}
%%%%%%%%%%%%%%%%%%%%%%% AFFILIATION %%%%%%%%%%%%
\vspace{0.2cm}
\begin{center}
${}^{a} $ {\small \it Faculty of Information Science and Technology, 
Osaka Institute of Technology,\\ Hirakata, Osaka 573-0196, Japan}\\[.2cm]
${}^{b}$ {\small \it Department of Physics and R-GIRO, Ritsumeikan University,
Kusatsu, Shiga, 525-8577, Japan}
\medskip
\vskip 10mm
\end{center}
\vskip 10mm
\begin{abstract}
We show in the framework of the Higgs triplet model that 
the lower bound of the neutrino mass is obtained 
if we can measure $\frac{\Gamma(\Delta^{--}\rightarrow ee)}{\Gamma(\Delta^{--}\rightarrow \mu\mu)}$ at the LHC.
\end{abstract}
\end{titlepage}
%%%%%%%%%%%%%%%%%%%%%%%%%%%%%%%%%%%%%%%%%%%%%%%%

Neutrino phenomena have been the guiding force for physics beyond the standard model (SM) for the last two decades.
Presently, observations of the Majorana nature of neutrinos and the absolute value of the neutrino mass would be strong evidence of particle physics.
The Higgs triplet model (HTM) \cite{HTM} is the simplest extension of the SM which invokes the new neutrino phenomena.
In a previous paper \cite{Fukuyama} we showed that the HTM enables us to detect the Majorana property by the precise measurement of the usual muon decay.
The interference term in muon decay due to the Majorana property was first discussed in \cite{Kotani}, but the value was far beyond the present upper bound.
On the contrary, the HTM gives a rather marginal value to the present precision order.
In the present paper we discuss another urgent problem of the absolute value of the neutrino mass that may be solved 
in the observations at the CERN LHC if the HTM works well. 

The neutrino-Higgs coupling in the HTM is given by
\begin{align}
{\mathcal L}_\text{HTM}^{} &= \overline{L^c}h_M\,i\tau_2\Delta
L+\text{H. c.}
\end{align}
Here neutrinos are required to be Majorana particles. The symmetric matrix
$h_M^{}$ is the coupling strength and $\tau_i(i=1$--$3)$ denote 
the Pauli matrices.
The triplet Higgs boson field with hypercharge $Y=2$ 
can be parametrized by
\begin{align}
\Delta=\begin{pmatrix}\Delta^+/\sqrt2&\Delta^{++}\\
\frac{v_\Delta^{}}{\sqrt2}+\Delta^0&-\Delta^+/\sqrt2\end{pmatrix},
\end{align}
where $v_\Delta^{}$ is the vacuum expectation value of 
the triplet Higgs boson. 
Mass eigenvalues of neutrinos are determined by diagonalization of
$m_\nu^{}=\sqrt2h_M^{}v_\Delta^{}$. 
There is a tree-level contribution to the electroweak $\rho$ parameter 
from the triplet vacuum expectation value as 
$\rho\thickapprox 1-2v_\Delta^2/v^2$.
The CERN LEP precision results can give an upper limit $v_\Delta^{}\lesssim 5$ GeV. 
There is no stringent bound from the quark sector on triplet Higgs bosons 
because they do not couple to quarks. 

The Yukawa interaction of the singly and the doubly charged Higgs boson
is written as
\begin{align}
{\mathcal L}_{\Delta}^{} =& 
-\sqrt2(h_M^\dag U)_{\ell i}\overline{\ell_L^{}}N_i^c\Delta^-\nonumber \\
&-(h_M^\dag)_{\ell\ell'}\overline{\ell_L^{}}{{\ell'}_L}^c\Delta^{--}
+\text{H.c.},
\end{align}
where 
\be
h_M^{}=Um_\nu^\text{diag}U^T/(\sqrt2v_\Delta^{})\equiv \left<m_\nu\right>_{ab}/(\sqrt2v_\Delta^{}), 
\ee
and 
$N_i$ represent Majorana neutrinos which satisfy the conditions $N_i=N_i^c=C\overline{N_i}^T$. 
The most stringent constraint on the triplet Yukawa coupling 
comes from $\mu\to ee{\bar e}$ through the tree-level contribution due to 
the doubly charged Higgs boson~\cite{Mu3eExp}. 
Thus the peculiar properties of the HTM appear in the processes of the doubly charged Higgs.
Among them, we have a sizable cross section of $\Delta^{++}\rightarrow l_al_b$ for $m_\Delta=O(100)$ GeV, and the rate is given by
\be
\Gamma(\Delta^{++}\rightarrow l_a^+l_b^+)=\frac{1}{4\pi(1+f)}|h_{ab}|^2m_{\Delta^{++}},
\ee
where $f=1(0)$ for $a=b$ ($a\neq b$). 

Searching for the neutrino mass in the HTM at the LHC was discussed in \cite{Hang}, where the event numbers were estimated at the scheduled energy and luminosity.
Unfortunately, the LHC was forced to lower the energy scale to $10$ TeV and the luminosity  to $10^{33}$ cm$^{-2}$s$^{-1}$.
The cross section for $pp\rightarrow \Delta^{++}\Delta^{--}$,
\bea
  99 &&\mbox{fb  for}~ m_{\Delta^{++}} = 200\mbox{GeV}\nonumber\\
 5.9 &&\mbox{fb  for}~ m_{\Delta^{++}} = 400\mbox{GeV},
\eea
for $\sqrt {s} = 14$TeV is reduced to
\bea
  53 &&\mbox{fb  for}~ m_{\Delta^{++}} = 200\mbox{GeV}\nonumber\\
 2.6 &&\mbox{fb  for}~ m_{\Delta^{++}} = 400\mbox{GeV},
\eea
for $\sqrt{s} = 10$TeV \cite{Sugiyama}.

The first 100-day run at low luminosity resulted in the integrated luminosity $10$ fb$^{-1}$.
So we may have a sizable number of events even in this case, though the final number of $ll$ events depends on Br($\Delta\rightarrow ll$).

In order to circumvent the ambiguous factors $m_\Delta$ and $v_\Delta$, let us consider the ratio
\be
\frac{\Gamma(\Delta^{--}\rightarrow ee)}{\Gamma(\Delta^{--}\rightarrow \mu\mu)}
=\left|\frac{\left<m_\nu\right>_{ee}}{\left<m_\nu\right>_{\mu\mu}}\right|^2.
\label{ratio}
\ee
Here the averaged masses are defined by 
\bea
\left<m_\nu\right>_{e e}&=&U_{e1}^2m_1+U_{e2}^2m_2+U_{e3}^2m_3,\\
\left<m_\nu\right>_{\mu \mu}&=&U_{\mu 1}^2m_1+U_{\mu 2}^2m_2+U_{\mu 3}^2m_3.
\eea
We have defined the averaged mass without an absolute symbol unlike the conventional one since $h_M$ is symmetric and complex matrix in general.
We know that the Maki-Nakagawa-Sakata-Pontecorv lepton mixing matrix $U$ is well approximated by the tribimaximal matrix \cite{tri}, and we assume here

\be
U=
\left(
        \begin{array}{ccc}
         \sqrt{\frac{2}{3}}  & \sqrt{\frac{1}{3}}e^{i\beta} &  0 \\
        -\sqrt{\frac{1}{6}}e^{-i\beta} & \sqrt{\frac{1}{3}} & -\sqrt{\frac{1}{2}}e^{i(\rho -\beta)}\\
         -\sqrt{\frac{1}{6}}e^{-i\rho}  & \sqrt{\frac{1}{3}}e^{-i(\rho-\beta)} & \sqrt{\frac{1}{2}}
        \end{array}
\right),
\ee
which is supplemented with the Majorana phases  $\rho$ and $\beta$.

For the normal hierarchy case, the neutrino mass is given by
\be
m_1=m_0,~~
m_2=\sqrt{m_0^2 + \Delta m_{sol}^2},~~
m_3=\sqrt{m_0^2 + \Delta m_{sol}^2 + \Delta m_{atm}^2},
\ee
where the smallest neutrino mass is denoted as $m_0$.
We adopt the center value of the neutrino mass squared difference~\cite{PDG} as
\be
\Delta m_{sol}^2=(8.0\pm 0.3)\times 10^{-5}~\text{eV}^2,~~\Delta m_{atm}^2=(1.9 - 3.0)\times 10^{-3}~\text{eV}^2
\ee

In this case, the branching ratio defined by Eq.(8) is a function of three parameters, 
namely, the smallest neutrino mass $m_0$, and the CP violating Majorana phases $\beta$ and $\rho$.
Therefore, a mesurement of the branching ratio at the LHC will lead us to a constraint among these three parameters.
That is, given the ratio of \bref{ratio}, the lower bound of $m_0$ is obtained as the minimum values of $m_0$ 
when we make $\beta$ and $\rho$ run over all possible values. This behavior is presented in Fig.~1. 
Thus the measurement of the ratio at the LHC will fix the lower bound of the neutrino mass $m_0$.
This is an interesting feature of the HTM.

For the inverse hierarchy case, the neutrino mass is given by
\be
m_3=m_0,~~
m_2=\sqrt{m_0^2+\Delta m_{atm}^2},~~
m_1=\sqrt{m_0^2+\Delta m_{atm}^2-\Delta m_{sol}^2},
\ee
where the smallest neutrino mass is denoted as $m_0$, as before.

In this case, the branching ratio defined by Eq.(8) is also a function of three parameters, 
$m_0$ and CP violating Majorana phases $\beta$ and $\rho$.
Therefore, a lower bound for the neutrino mass $m_0$ can be obtained in the same way as in the normal hierarchy case. 
The result is presented in Fig.~2. 

Thus the measurement of \bref{ratio} at the LHC leads us to the lower bound of the neutrino mass.
An interesting problem still remains---how the non-zero $\theta_{13}$ affects the result of the present paper, 
which will be discussed in a subsequent work.

After completing this paper, we learned that analogous arguments were discussed by Kadastik {\it et al.}~\cite{Kadastik}. 
They considered many channels, whereas we have studied in depth the most manageable $ee$ and $\mu\mu$ processes restrictively.
Indeed, the Majorana CP phases lead us to different views than \cite{Kadastik}, even in the same diagram.

\vspace{5mm}

{\bf Acknowledgments}~~~\\[2mm]
We are grateful to H.~Sugiyama, N.~Okamura, and K.~Tsumura for useful comments.
The work of T.~F. is supported in part by the Grant-in-Aid 
for Scientific Research from the Ministry of Education, 
Science and Culture of Japan (No. 20540282).

%%%%%%%%%%%%%%%%%%%%%%%%%%%%%%%%%%%%
\newpage

{\scalebox{0.9}{\includegraphics{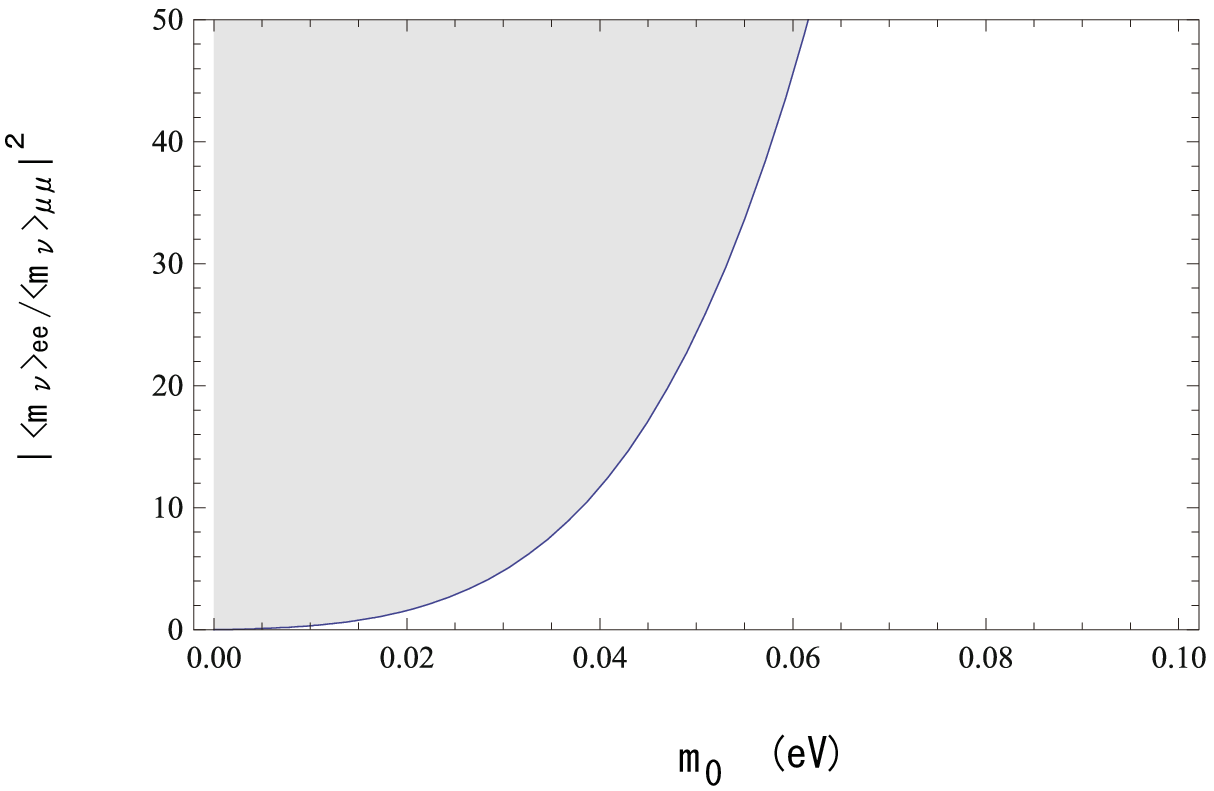}} }

\begin{quotation}
{\bf Fig.~1}  Behavior of $\frac{\Gamma(\Delta^{--}\rightarrow ee)}{\Gamma(\Delta^{--}\rightarrow \mu\mu)}
=\left|\frac{\left<m_\nu\right>_{ee}}{\left<m_\nu\right>_{\mu\mu}}\right|^2$
versus $m_0$ in the normal hierarchy case for the neutrino mass. 
The nonshaded area is allowed, and is obtained by running over all possible values of $\beta$ and $\rho$.
\end{quotation}

\vspace{5mm}

{\scalebox{0.9}{\includegraphics{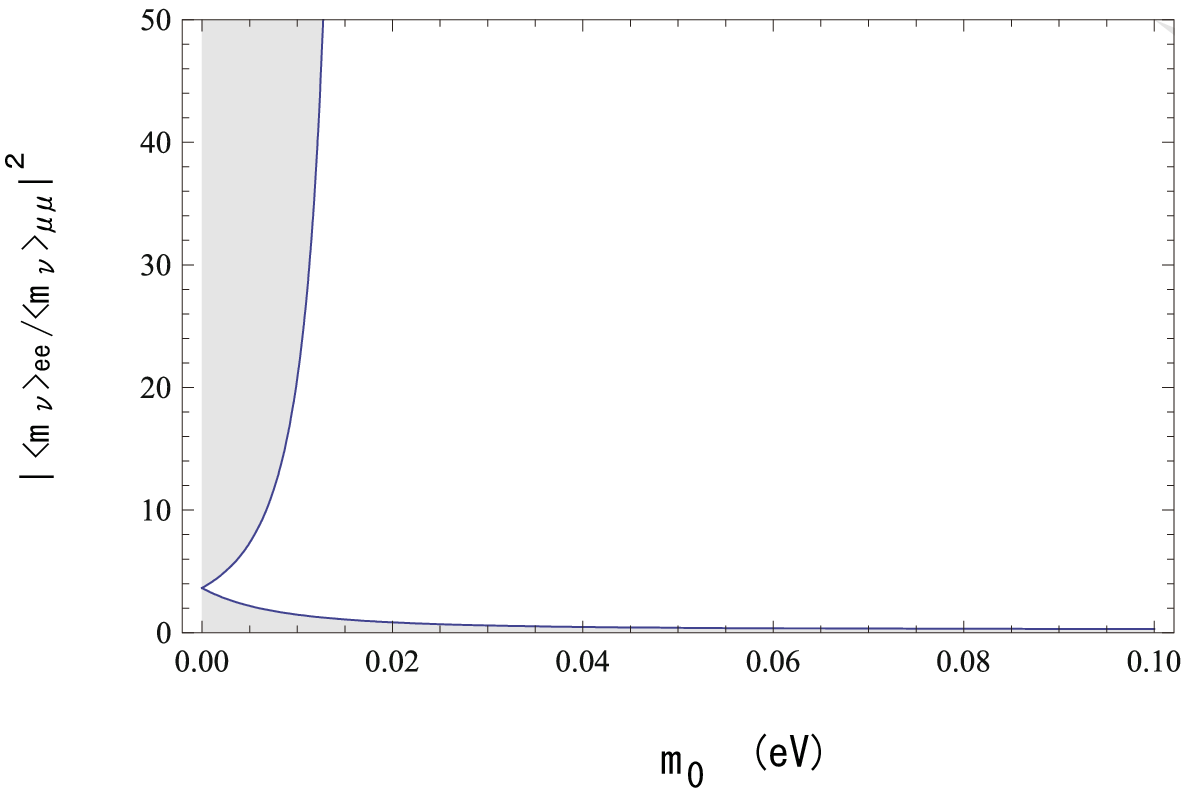}} }

\begin{quotation}
{\bf Fig.~2}  Behavior of $\frac{\Gamma(\Delta^{--}\rightarrow ee)}{\Gamma(\Delta^{--}\rightarrow \mu\mu)}
=\left|\frac{\left<m_\nu\right>_{ee}}{\left<m_\nu\right>_{\mu\mu}}\right|^2$
versus $m_0$ in the inverse hierarchy case for the neutrino mass. 
The nonshaded area is allowed, and is obtained by running over all possible values of $\beta$ and $\rho$. 
We cannot find the lower bound around the ratio=4.
\end{quotation}

\end{document}